\documentclass[12pt]{article}

\makeatletter
\renewcommand\section{\@startsection {section}{1}{\z@}%
                                   {-3.5ex \@plus -1ex \@minus -.2ex}
                                   {2.3ex \@plus.2ex}%
                                   {\normalfont\large\bfseries}}
\renewcommand\subsection{\@startsection{subsection}{2}{\z@}%
                                     {-3.25ex\@plus -1ex \@minus -.2ex}%
                                     {1.5ex \@plus .2ex}%
                                     {\normalfont\bfseries}}
\makeatother
\newcommand{\beq}{\begin{equation}}
\newcommand{\eeq}{\end{equation}}
\newcommand{\ber}{\begin{array}}
\newcommand{\eer}{\end{array}}
\newcommand{\D}{{\cal D}}

\newcommand{\dtwo}{d^{\hspace{1pt}2}\hspace{-1pt}}
\newcommand{\del}{\partial}

\newcommand{\dsty}{\displaystyle}

\newcommand{\te}{\theta}

\newcommand{\de}{\delta}
\newcommand{\ds}{\dtwo\sigma}

\newcommand{\eps}{\varepsilon}

\newcommand{\ena}{\end{eqnarray}}
\newcommand{\beqa}{\begin{eqnarray}}
\newcommand{\eeqa}{\end{eqnarray}}
\newcommand{\bea}{\begin{eqnarray}}
\newcommand{\eea}{\end{eqnarray}}

\newcommand{\be}{\begin{equation}}
\newcommand{\ee}{\end{equation}}

\begin{document}
\begin{titlepage}
\begin{flushright}
\phantom{arXiv:yymm.nnnn}
\end{flushright}
\vfill
\begin{center}
{\Large\bf Quantum backreaction in string theory}    \\
\vskip 15mm
Oleg Evnin
\vskip 10mm
{\em Department of Physics, Faculty of Science, Chulalongkorn University,\\
Phayathai Rd., Bangkok 10330, Thailand}
\vskip 5mm
{\em Institute of Theoretical Physics, Academia Sinica\\
Zh\=onggu\=anc\=un d\=ongl\`u 55, Beijing 100190, China}
\vskip 5mm
{\small\noindent  {\tt oleg.evnin@gmail.com}}
\end{center}
\vfill

\begin{center}
{\bf ABSTRACT}\vspace{3mm}
\end{center}

There are situations in string theory when a finite number of string quanta
induce a significant backreaction upon the background and render the perturbation theory
infrared-divergent. The simplest example is D0-brane recoil under an impact by closed strings.
A more physically interesting case is backreaction on the evolution of a totally
compact universe due to closed string gas. Such situations necessitate qualitative amendments to the traditional formulation of string theory in a fixed classical background. In this contribution to the proceedings of the XVII European Workshop on String Theory in Padua, I review solved problems and current investigations in relation to this kind of {\it quantum backreaction} effects.

\vfill

\end{titlepage}

\section{Introduction}

In the conventional formulation of perturbative string theory, one starts with a classical space-time background and considers how strings propagate in it. A subset of string modes is then found to describe deformations of the background. String theory has often been criticized by its opponents for this lack of (explicit) background independence, but it may well be that quantum gravity theories must be formulated with a specification of the background (at least asymptotically).

Be it as it may, there are situations when the conventional separation of the classical background and stringy excitations simply cannot be applied. This typically happens when the background is ``small'' in some sense, such as the number of non-compact dimensions or spatial extent -- the precise meaning of these words will become clear as our discussion progresses. In such situations, a finite number of string quanta exert a substantial backreaction on the background, making the original background specification meaningless (as it is modified significantly and in different ways by each string propagation process).

A simple example is given by D0-brane recoil. The usual CFT description of D0-branes is formulated in the background of straight D0-brane worldlines. The problem is that, once hit by closed strings, the D0-brane starts to move. Unlike higher-dimensional branes, a recoiling D0-brane will get displaced infinitely far away from its initial position, deviating infinitely strongly from the original classical background in which we formulated the theory.\footnote{Note that, even though the recoil of point-like objects is an elementary process, and its role in perturbative dynamics of solitons and D0-branes is widely appreciated, it is less well-known that there is a related recoil effect for string-like topological defects and D1-branes requiring special attention \cite{localrecoil}.} Furthermore, since the strings that hit the D0-brane are quantum objects and do not have a definite position in space, one does not expect the recoiling D0-brane to have a definite position in space either. Hence replacing the background of a straight D0-brane worldline by any other classical trajectory is not expected to solve the problem.

The recoil process is physically trivial (though completely non-trivial to implement in string perturbation theory), but here is a more interesting example: consider strings in a flat toroidally compactified space. If the number of compactified dimensions is not too large, the average energy density induced by putting a finite number of strings in this space is zero, so one does not expect significant backreaction. However, if there are no non-compact spatial dimensions\footnote{Again, one should also expect subtleties when there is only one non-compact spatial dimension.}, a finite number of strings will induce a finite energy density, making the space-time evolve far away from the original flat background. Since closed strings are quantum objects, they do not have definite positions in space, and one should not expect their effect on the geometry to amount to simply replacing the original flat background by another classical background.

The physics of the above-mentioned processes is rather transparent, but their implementation in string perturbation theory is extremely subtle. It is a common lore that expanding around a wrong background leads to infrared divergences in perturbation theory, and string theory is no exception. The divergences are deferred to the first subleading order in string coupling in both of the above-mentioned cases (because, in the perturbative regime, D0-branes are very massive compared to string scale, and the Planck scale controlling gravitational backreaction is much greater than the string scale as well). However all perturbative corrections starting from the next-to-leading order are infinite because of the infrared divergences, making the standard string perturbation theory {\it ill-defined}. (Strings in toroidal spaces have been a subject of many interesting phenomenological speculations, including the Brandenberger-Vafa scenario \cite{bv}, and it is regrettable that a fundamental formulation of string theory in such backgrounds has not been given until now.)

It has been long understood that solutions to infrared troubles of string perturbation theory should lie in an appropriate implementation of the Fischler-Susskind mechanism \cite{pt,fpr} (these papers were dealing with the recoil case). Namely, one needs to modify conventional string theory by unusual perturbative corrections that induce a divergence on the worldsheet of the leading genus (disk, sphere) that would cancel the infrared divergence on the subleading genus (annulus, torus) indicative of backreaction (with a hope that the construction could extend to higher orders of perturbation theory). However, without additional systematic framework, it is difficult to specify correct deformations of the conventional string theory that do this job. In particular, the deformations proposed in \cite{pt,fpr} are unsatisfactory in a number of ways \cite{d0,thesis}.

One systematic strategy for dealing with quantum backreaction effects is to choose a set of infrared modes of the background (position of the D0-brane, scale factor of the cosmological space) and quantize them explicitly. A criterion by which this approach should be judged is whether it leads to a successful realization of the Fischler-Susskind mechanism, automatically cancelling the infrared divergences indicative of backreaction. This program has been implemented for the D0-brane recoil case in \cite{d0,thesis}, and investigations are currently underway to implement it for the cosmological case.

One may hear objections to explicit quantization of the infrared modes of the background based on the perception that such modes are already included among the string excitations. There is a vicious circle in such reasoning. The intuition that strings describe excitations of the classical background stems from an explicit analysis of conventional perturbative string theory in backgrounds in which it is well-defined. For our present settings, conventional perturbative string theory is plainly divergent, and there are no grounds for assuming ad hoc that string excitations describe this or that. One needs to formulate a finite consistent theory first, and then look for its qualitative interpretations. Explicit quantization of a subset of the infrared modes of the background provides such a theory for the D0-brane case (and, hopefully, for the cosmological case as well).

In what follows, I briefly review the results of \cite{d0,thesis} regarding D0-brane recoil and some preliminary investigations in relation to the case of cosmological backreaction from a gas of quantum strings.

\section{D-particle recoil}

\subsection{The annular divergence}

We shall start by exposing explicitly the problem that forces one to reconsider the standard definition of perturbative string theory in the background of a recoiling D0-brane, namely, the infrared divergence at next-to-leading order in string coupling.
The divergence comes from annulus worldsheets developing a long, thin strip. Divergences from degenerating Riemann surfaces can be analyzed using Polchinski's plumbing fixture construction \cite{polch9, polchinski-fischler-susskind}, which relates the divergences to amplitudes evaluated on a lower genus Riemann surface. In particular, the annulus amplitude with an insertion of vertex operators $V^{(1)},\cdots,V^{(n)}$ (in the interior) can be expressed through disk amplitudes with additional operator insertions at the boundary:
\beq
\left\langle V^{(1)}\cdots V^{(n)}\right>_{annulus}=\sum\limits_\alpha \int \frac{dq}q\, q^{h_\alpha-1} \int d\te d\te' \left\langle V_\alpha(\te)V_\alpha(\te')V^{(1)}\cdots V^{(n)}\right>_{disk},
\label{plumbing}
\eeq
where the summation extends over a compete set of local operators $V_\alpha(\te)$ with conformal weights $h_\alpha$, and $q$ is the gluing parameter, which can be related to the annular modulus. ($\te$ parametrizes the boundary of the disk.) The divergence in the integral over $q$ coming from the region $q\approx0$ (i.e., from an annulus developing a thin strip) will be dominated by the terms with the smallest possible $h_\alpha$. 

Neglecting the tachyon divergence, which is a pathology peculiar to the case of the bosonic string and should not be present in more realistic settings, we consider the following operators with conformal weights $h=1+\alpha'\omega^2$:
$V^i(\te)= \del_n X^i(\te)\exp\left[i\omega X^0(\te)\right]$. These operators correspond to massless open string states (representing translations of the D0-brane in the $i$'th Dirichlet direction). For small values of $q$ (which is the region we are interested in), only small values of $\omega$ will contribute to the integral. Hence, the annular divergence takes the following form:
\bea
\left\langle V^{(1)}\cdots V^{(n)}\right>_{annulus}^{(div)}\sim\int\limits_0^1 dq\int\limits_{-\infty}^\infty  d\omega\, q^{-1+\alpha'\omega^2}\int d\te d\te' \left\langle V^i(\te,\omega)V^i(\te',\omega)V^{(1)}\cdots V^{(n)}\right>_{disk}\nonumber\\
\phantom{.}\hspace{-3cm}\sim P^2\left\langle V^{(1)}\cdots V^{(n)}\right>_{disk}\int\limits_0^1 dq\int\limits_{-\infty}^\infty d\omega\, q^{-1+\alpha'\omega^2}\sim P^2\left\langle V^{(1)}\cdots V^{(n)}\right>_{disk}\int\limits_0^1 {dq\over q\,(-\log q)^{1/2}}\ ,
\label{annulus}
\eea
where we have taken into account the fact that the operator $\int \del_n X^i(\te)d\te$ merely shifts the position of the D0-brane; inserting it into any amplitude amounts to multiplication by the total (Dirichlet) momentum $P$ transferred by the closed strings to the D0-brane during scattering. Introducing a cut-off $\eps$ on the lower bound of the $q$-integral in (\ref{annulus}) reveals a  $\sqrt{|\log\eps|}$ divergence that is indicative of recoil:
\beq
\left\langle V^{(1)}\cdots V^{(n)}\right>_{annulus}^{(div)}\sim P^2\left<V^{(1)}\cdots V^{(n)}\right>_{disk}\sqrt{|\log\eps|}.
\label{annulusD0}
\eeq
The fact that the divergence is proportional to the square of the transferred momentum highlights its relation to recoil (the divergence vanishes when no momentum is transferred). The overall normalization can be determined with additional effort (see \cite{d0}).

\subsection{Worldline formalism}

The annular divergence we have displayed signifies a breakdown of traditional perturbative string theory in the background of a recoiling D0-brane. A systematic, physically intuitive modification of the formalism is to introduce a fully dynamical trajectory of the D0-brane described by a path integration variable $f^\mu(t)$. Such a formalism was proposed in \cite{hk}, substantially reorganized and strengthened in \cite{worldline} and successfully applied to the recoil problem in \cite{d0,thesis}.

The main object of interest is the amplitude for a D0-brane to move from the point $x_1^\mu$ to the point $x_2^\mu$ while absorbing/emitting closed strings carrying momenta $k_n$:
\beq
\ber{l}
\dsty G(x_1,x_2|\,k_n)=\sum\frac{\left(g_s\right)^\chi}{V_\chi} \int \D f\, \D t\,\D X\,\de\left(X^\mu(\te)-f^\mu(t(\te))\right)e^{-S_{D}(f)-S_{st}(X)} \prod\left\{g_s V^{(n)}(k_n)\right\},
\eer
\label{master}
\eeq
where $S_{st}$ is the standard conformal gauge action
($4\pi\alpha'S_{st}=\int\ds\nabla X_\mu\nabla X^\mu$),
the integration with respect to $f_\mu$ extends over all the inequivalent (unrelated by diffeomorphisms) curves starting at $x_1$ and ending at $x_2$, the boundary of the worldsheet is parametrized by $\te$, and $t(\te)$ describes how this boundary is mapped onto the D0-brane worldline. The sum is over all the topologies of the worldsheets (not necessarily connected, but without any disconnected vacuum parts) and $\chi$ is the Euler number. $V_\chi$ is the conformal Killing volume (the negative regularized value of \cite{volume} should be used for the disk). The fully integrated form of the vertex operators is implied. The integration over moduli
of the worldsheet is suppressed, since we shall be mostly working with worldsheets of disk topology. The scattering amplitude can be deduced from (\ref{master}) by means of the standard reduction formula:
\beq
\left<p_1|p_2\right>_{k_n}=\lim\limits_{p_1^2,p_2^2\to -M^2} \left(p_1^2+M^2\right)\left(p_2^2+M^2\right)\int dx_1dx_2 e^{ip_1x_1} e^{ip_2x_2}G(x_1,x_2|\,k_1,\cdots,k_m), 
\label{reduct}
\eeq
where $M$ is the D0-brane mass. The choice of the worldline action $S_D$ is subtle (see \cite{d0,thesis,hk,worldline}) but for the lowest order derivations it suffices to consider the simplest free-particle action (length of the worldline multiplied by the D0 mass): $S_D[f(t)]=MT$.

\subsection{Divergence cancellation}

It is essential to realize that the path integral (\ref{master}) contains contributions from curved worldlines, and curved D0-brane worldlines will lead to ultraviolet divergences in the worldsheet path integral (and related Weyl anomaly). The situation is completely analogous to a more familiar case: string propagation in a curved space-time \cite{gsw,ct}. Note also that the D0-brane is heavy and the typical curvature of its worldline is small. Hence the divergences (on the disk, for the purposes of our considerations) will gain an additional power of $g_s$, which will put them in the right order of string coupling to cancel the annulus divergence (\ref{annulusD0}).

The intuitive expectation that making D0-brane worldlines dynamical automatically cancels the annulus divergence that invalidates conventional string perturbation theory proves to hold upon evaluation of the path integral (\ref{master}). The reader should consult \cite{d0,thesis} for detailed derivations. Schematically, one finds the following relation between the divergent part of the next-to-leading order contribution to the scattering amplitude (\ref{reduct}) on the disk $\left<p_2|p_1\right>^{(1;div)}$, and the (finite) leading order contribution to the same amplitude $\left<p_2|p_1\right>^{(0)}$:
\beq
\left<p_2|p_1\right>^{(1;div)}\sim\frac{P^2}{M}\left<p_2|p_1\right>^{(0)}\int d\omega\, G_{kick}(\omega)e^{2\alpha'\omega^2\log\de}\sim\sqrt{\alpha'|\log\de|}\,\,\frac{P^2}{M}\left<p_2|p_1\right>^{(0)},
\label{diskdiv}
\eeq
where $G_{kick}(\omega)\sim\frac{1}{(\omega+i0)^2}+\frac{1}{(\omega-i0)^2}$ is the Fourier transform of the Green function of free motion on a line $G_{kick}(t)=|t|/2$, and $\de$ is a small-distance worldsheet cut-off. One can recognize a structure similar to (\ref{annulusD0}) and matching momenta dependences, and upon closer inspection (with an appropriate geometrically-inspired identification of minimal distance cut-off on the disk and modulus cut-off on the annulus) $\left<p_2|p_1\right>^{(1;div)}$ exactly cancels the divergent contribution to the same process coming from the annulus, thus successfully implementing the Fischler-Susskind mechanism at next-to-leading order in string coupling.

It should be noted for the sake of future discussion that (\ref{diskdiv}) manifestly contains contributions of different orders in $\alpha'$ whereas in familiar computations of worldsheet divergences in fixed classical backgrounds \cite{gsw,ct} one is accustomed to performing $\alpha'$-expansions. It is thus obvious that, if following a more conventional route, one would need to perform a resummation of leading logarithmic divergences in order to arrive at (\ref{diskdiv}). (Such a resummation has not been performed in \cite{hk}, which is one of the reasons why the implementation of the Fischler-Susskind mechanism has not been elucidated in that publication.)

\section{Backreaction in a totally compact space}

\subsection{The torus divergence}

We shall now turn to the case of strings in a totally compact cosmological space. Just like for the case of D0-brane recoil, it is important to see how na\"\i ve attempts to formulate string perturbation theory fail.

For example, one can consider strings on a flat space-time with all spatial directions compactified on circles ($R\times T^{25}$ for bosonic string theory). One often hears that massless string states describe deformations of the background (and this intuition is perfectly sound for non-compact Minkowski space backgrounds), however for this case when quantum backreaction occurs, conventional perturbative string theory is plainly divergent and fails to account for the cosmological dynamics of the compact background space.

Just like in the D0-brane recoil case, the first divergence comes at next-to-leading order in string coupling, which corresponds to a torus worldsheet. More specifically, the divergence arises from the limit when the torus develops a long thin handle (described in this case by a single complex modular parameter $q$). As usual, via the plumbing fixture construction \cite{polch9, polchinski-fischler-susskind}, the divergence can be expressed through sphere amplitudes with two additional vertex operator insertions corresponding to massless string states carrying zero momentum in all compact directions, multiplied by propagator factors
\beq
\int d\omega\int_{|q|<1}\frac{d^2q}{q\bar q}(q\bar q)^{\alpha'\omega^2/4}\sim\int \frac{d\kappa}{\kappa(-\log\kappa)^{1/2}},
\label{cosmdiv}
\eeq
where $\kappa=(q\bar q)^{1/2}$. Cutting off the lower bound of the $\kappa$-integration at $\kappa=\eps$ reveals, just as it did in the recoil case, a $\sqrt{|\log\eps|}$ divergence.

\subsection{Strings in a quantum cosmological space}

Motivated by the success of explicit quantization of a set of infrared modes of the background in the case of D0-brane recoil, we can try to devise a similar treatment for the cosmological case. Since the divergence (\ref{cosmdiv}) comes specifically from massless string modes carrying zero momentum in compact spatial directions, it is natural to consider a class of backgrounds made of precisely such modes:
\beq
ds^2=-N^2(t)dt^2+a^2(t)dx^idx^i
\label{metric}
\eeq
(supplemented by a time-dependent dilaton $\phi(t)$ and, if necessary, time-dependent 2-form field). One would then have to perform path integration over $a(t)$ and $\phi(t)$ in addition to the conventional string worldsheet path integral. The hope is that such a path integral will lead to an automatic cancellation of infrared divergences (as it did in the recoil case).

Quantization of a restricted set of gravitational modes in a compact cosmological space has been extensively studied under the name of ``minisuperspace'' models. It is known that quantization of these models (as well as defining physical observables after the quantization) is very ambiguous and no unique quantization prescription has been devised (for an extensive review, see \cite{isham}). Nevertheless, a pragmatic approach would be to adopt a particular prescription for the quantization of the infrared background modes, supplement it by the worldsheet path integral and examine worldsheet divergence cancellation in the spirit of the Fischler-Susskind mechanism.

A conservative quantization prescription (described, in particular, in \cite{marolf}) is to apply a Faddeev-Popov treatment to the time-repar
Such a treatment naturally eliminates all time-dependence in the usual transition amplitudes defined by path integrals, and the resulting time-independent ``transition amplitudes'' are treated as scalar products of state vectors and declared to be the fundamental observables of the theory. Such scalar products can be schematically represented as
\beq
\langle a_2,\phi_2|a_1,\phi_1\rangle=\int\limits_{-\infty}^{\infty}dT\int \D a(t) \D\phi(t) e^{-S_{a\phi}},
\label{Mrolf}
\eeq
where $S_{a\phi}$ is the Einstein-Hilbert action in the background (\ref{metric}), and the boundary conditions in the path integral over $a(t)$ and $\phi(t)$ are specified to be $(a_1,\phi_1)$ at the moment 0 and $(a_2,\phi_2)$ at the moment $T$, with a subsequent integration over $T$ (arising from the time-reparametrization invariance).

One can try to extend the minisuperspace path integral by the full set of string modes defining the following quantities:
\beq
\langle a_2,\phi_2|a_1,\phi_1\rangle_{V}=\int\limits_{-\infty}^{\infty}dT\int \D a(t) \D\phi(t) e^{-S_{a\phi}}\int \D X e^{-S_X} \{g_s\prod V^{(n)}\},
\label{Mrolf_string}
\eeq
where $S_X$ is the string worldsheet action in the background (\ref{metric}) and a product of string vertex operators $V^{(n)}$ has been inserted in the path-integral. The question is then whether divergent contributions to (\ref{Mrolf_string}) arising from sphere worldsheets in curved space-time backgrounds and the modular integration divergence for torus worldsheets will cancel each other as they did for the D0-brane recoil problem.

Qualitatively, the theory defined by (\ref{Mrolf_string}) can be viewed in two different ways. We arrived at it by trying to devise a modification of standard perturbative string theory capable of accounting for backreaction from string gas in a totally compact space and introducing explicit path integration over a set of infrared background modes (in addition to the worldsheet path integration). The theory could equally well arise in an attempt to devise a stringy ultraviolet completion for minisuperspace models.

The main technical problem that remains is to show that the Fischler-Susskind divergence cancellation in the context of (\ref{Mrolf_string}) works as intended. (One is accustomed to evaluating wordsheet divergences in curved space-time backgrounds via $\alpha'$-expansion \cite{gsw,ct}. However, the D0-brane recoil considerations teach us that resummation of leading divergences to all orders in $\alpha'$ is likely to be necessary, before the divergence cancellation becomes manifest.) Future work in this direction will be reported elsewhere \cite{future}.

\section{Acknowledgments}

I would like to thank Ben Craps, Anatoly Konechny and Shin Nakamura for collaboration on various portions of the research presented here, and the organizers of the XVII European Workshop on String Theory in Padua for arranging the meeting and the publication of these proceedings. This research has been supported by grants from the Chinese Academy of Sciences, National Natural Science Foundation of China and Ratchdapisek Sompote Endowment Fund.

\end{document}